\documentclass[twocolumn,aps,prb,superscriptaddress]{revtex4}
\usepackage{amssymb}
\usepackage{graphicx}
\usepackage{dcolumn}
\usepackage{bm}
\usepackage{amsmath}
\usepackage{ulem}
\usepackage[colorlinks,linkcolor=magenta,citecolor=blue,urlcolor=blue]{hyperref}
\usepackage{verbatim}

\begin{document}

\title{Exact solution of single impurity problem in non-reciprocal lattices: impurity induced size-dependent non-Hermitian skin effect}
\author{Yanxia Liu}
\affiliation{Beijing National Laboratory for Condensed Matter Physics, Institute of
Physics, Chinese Academy of Sciences, Beijing 100190, China}
\author{Yumeng Zeng}
\affiliation{Beijing National Laboratory for Condensed Matter Physics, Institute of
Physics, Chinese Academy of Sciences, Beijing 100190, China}
\affiliation{School of Physical Sciences, University of Chinese Academy of Sciences,
Beijing 100049, China}
\author{Linhu Li}
\email{lilh56@mail.sysu.edu.cn}
\affiliation{Guangdong Provincial Key Laboratory of Quantum Metrology and Sensing $\&$ School of Physics and Astronomy, Sun Yat-Sen University (Zhuhai Campus), Zhuhai 519082, China}
\author{Shu Chen}
\email{schen@iphy.ac.cn}
\affiliation{Beijing National Laboratory for Condensed Matter Physics, Institute of
Physics, Chinese Academy of Sciences, Beijing 100190, China}
\affiliation{School of Physical Sciences, University of Chinese Academy of Sciences,
Beijing 100049, China}
\affiliation{Yangtze River Delta Physics Research Center, Liyang, Jiangsu 213300, China}

\begin{abstract}
Non-Hermitian non-reciprocal systems are known to be extremely sensitive to boundary conditions,
exhibiting diverse localizing behaviors and spectrum structures when translational invariance is locally broken, either by tuning the boundary coupling strength, or by introducing an effective boundary using impurities or defects.
In this work, we consider the single impurity problem in the Hatano-Nelson
model and the Su-Schreieffer-Heeger model, which can be exactly solved with the single impurity 
being treated as an effective boundary of the system. From our exact 
solutions for finite-size systems, we unveil that increasing the impurity 
strength can lead to a transition of the bulk states from non-skin states to skin states, 
accompanied by the change of the spectrum structure from an ellipse in the complex plane to a segment along the real axis.
These exact results indicate that the critical value of impurity 
strength is size-dependent, and increases exponentially with the lattice size when the impurity is strong or the system is large enough. 
Our exact solutions are also useful for determining the spectral topological transition in the concerned models.
\end{abstract}

\maketitle

\section{Introduction}
Boundary conditions play a pivotal role in determining the properties of a wide variety of physical systems,
ranging from the most fundamental problem of solving a single-particle Schr\"odinger equation, 
to more stimulating emergent phenomena such as the quantum Hall effect, 
where the quantized Hall conductance is associated with the number of topological states localized at the physical boundaries of the system.
In contrast, the bulk energy-spectra are usually expected to be insensitive to boundary perturbations, as their corresponding eigenstates are mostly distributed in the bulk of the system.
However, this picture generally fails when non-Hermiticity is introduced to the Hamiltonian, 
where the spectra under periodic and open boundary conditions (PBC and OBC) can dramatically diverge from each other \cite{higham2005spectra}. 
Physically, such a significant boundary effect can be understood with the non-Hermitian skin effect (NHSE), namely a majority of eigenstates are pumped to the boundaries by the non-reciprocity of the system under OBC \cite{Yao}.
To date, the NHSE has been extensively investigated in various systems, \cite{Yao,Yin,Alvarez,Gong,Kunst,Yao1,Jin,Lee,Kawabata,Lee1,WangZhong2019,Herviou,Kou,Rui,Xue,Longhi,JiangHui2019,Turker,Ueda,KZhang,KYokomizo,LeeCH,Okuma,HShen,Budich,Sato-PRB,Kunst-PRB,RChen,Budich-EPJD,CHLiu2020,CSE,directional_amplification1,directional_amplification2,Borgnia}
as it is known to be associated with many intriguing non-Hermitian phenomena, e.g. the breakdown of conventional bulk boundary correspondence \cite{Yao,Xiong}, the spectral point-gap topology \cite{Borgnia,KZhang,Okuma}, the critical NHSE \cite{CHLiu2020,CSE}, and the directional signal amplification \cite{directional_amplification1,directional_amplification2}.

Beyond the PBC and OBC, much effort has been made recently in exploring non-Hermitian systems with other types of boundary conditions.
It has been found that by tuning the strength of boundary hoppings away from both the PBC and OBC, 
a new type of so-called scale-free accumulating states emerges in a finite-size system,
and the NHSE becomes less stable against such boundary perturbations when increasing the system's size \cite{Linhu,guo2021exact}.
The continuous deformation between the PBC and OBC also provides more insight of the NHSE \cite{LeeCH,LeeCH-PRB2020},
and leads to a topological quantized response unique in non-Hermitian systems \cite{li2020quantized}.
On the other hand, strong impurities and defects effectively induce a boundary in a periodic system, and may act as OBC for boundary phenomena in either Hermitian and non-Hermitian systems \cite{Kane,LiuCH2,Slager2015,ShenSQ,Kimme,Lang2014,LiuYX}.

In this paper, we analytically study the single impurity problem in two representative 1D non-Hermitian models that exhibit the NHSE under OBC, namely the Hatano-Nelson (HN) model and the non-reciprocal Su-Schreieffer-Heeger (SSH) model.
In both cases, exact solutions are obtained with a single on-site impurity potential, whose strength varies from zero to infinity.
The bulk states are seen to go through a transition from non-skin states to skin states at certain critical values of the impurity strength, after which their eigenenergies become purely real.
In other words, a strong impurity behaves similarly as the OBC, except for the bound state localized at the impurity.
The transition value of the impurity strength is found to depend on the system's size and the non-reciprocity strength of the system, and also varies for different eigenstates.
We have also applied our results to study the spectral topology of these two models, and the topological transition points are accurately predicted by our exact solutions. 

The rest of the paper is organized as follows. 
In Sec. II, we present the exact solutions for the single-impurity problem in  the HN model, and analyze in details the transition for the bulk states from non-skin to skin states. 
In Sec. III, we study the single-impurity problem for the non-reciprocal SSH model, where exact solutions can be obtained by mapping it to the HN model.  
We then applies our exact solutions to identify the topological transition points of the spectral topology of the two models in Sec. IV.
Finally, a summary of our results is given in Sec. V.

\section{Single impurity problem for one-band 1D non-reciprocal lattice}
\begin{figure}[tbp]
\includegraphics[width=0.45\textwidth]{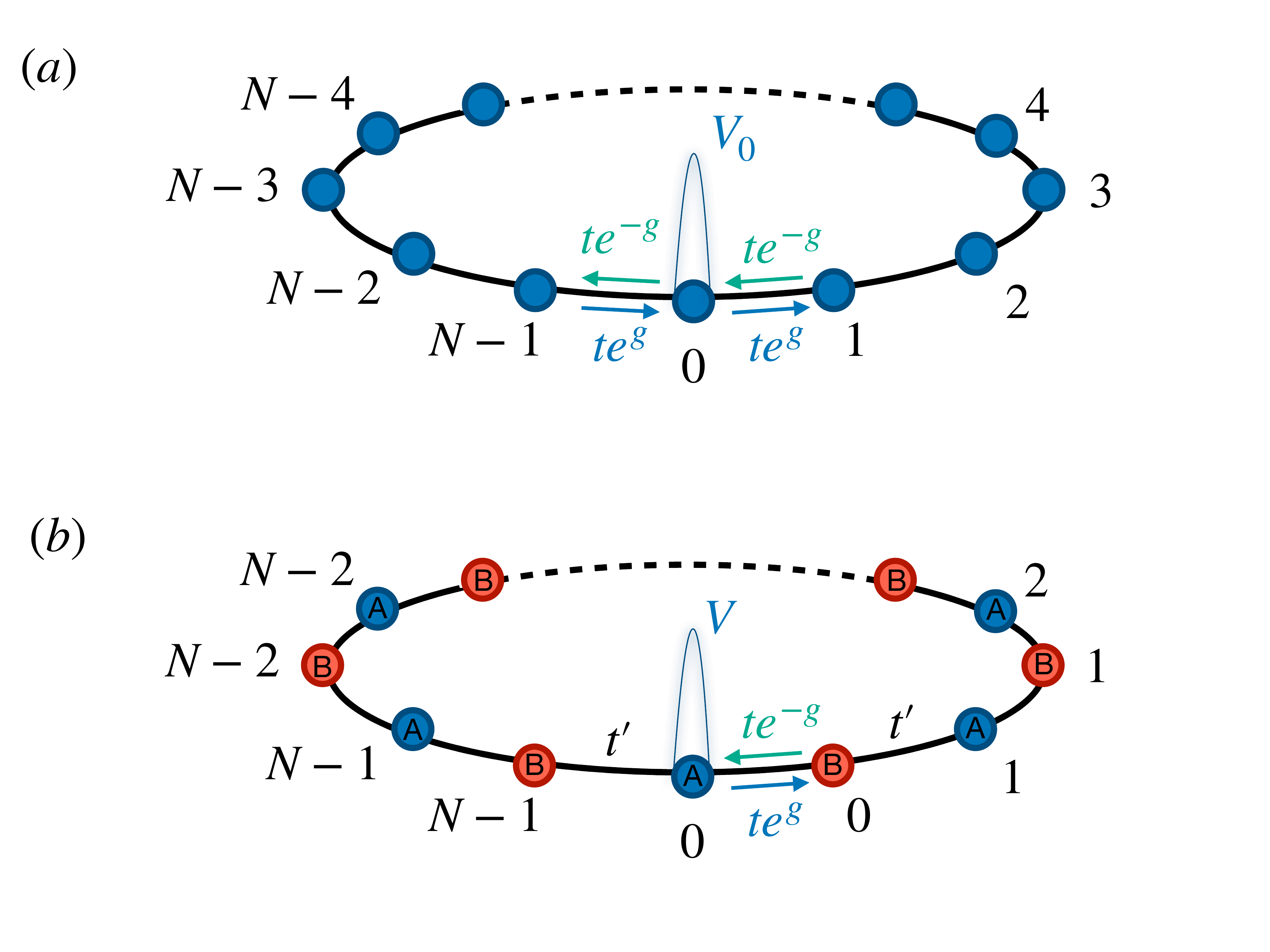}
\caption{(a) The Hatano-Nelson model with one impurity.
(b)The non-reciprocal SSH model with one impurity. }
\label{fig1}
\end{figure}

We consider a 1D HN chain \cite{HN_chain} with an impurity under the periodic boundary
condition [see Fig. \ref{fig1}(a)]. The Hamiltonian is given by
\begin{equation}
H_{\rm NH}=\sum_{n=0}^{N-1}\left( t_{L}\left\vert n\right\rangle \left\langle
n+1\right\vert +t_{R}\left\vert n+1\right\rangle \left\langle n\right\vert
\right)+V_0 \left\vert 0\right\rangle \left\langle 0\right\vert ,  \label{H0}
\end{equation}%
where $t_{R(L)}$ denotes the right (left)-hopping amplitude
which can be parameterized as $t_{L}=te^{-g}$ and $t_{R}=te^{g}$ with real $%
t $ and $g$. Here we set $t=1$ as the energy unit.
The asymmetry of hopping amplitudes $\left( g\neq 0\right) $ leads to the non-Hermiticity of the model. 
Nevertheless, the Hamiltonian is pseudo-Hermitian as it satisfies $H^{\dag}=PHP$ with
$P$ the parity operator. That is, the Hamiltonian $H$ becomes $H^\dag$ when exchanging sites $i$ and $N-i$ in Fig. \ref{fig1}(a).
Therefore the eigenenergies of $H$ are either real, or given by complex conjugated pairs.


The Schr\"odinger equation $H\left\vert \psi \right\rangle=\epsilon \left\vert \psi \right\rangle$
with the wave function $\left\vert \psi \right\rangle=\sum_n \psi_n \left\vert n\right\rangle$
can be written as a second order homogeneous linear difference equation
\begin{equation}
e^{g}\psi_{n-1} +e^{-g}\psi _{n+1} = \epsilon \psi _{n},  \label{eigeneq}
\end{equation}%
with boundary conditions
\begin{equation}
e^{g}\psi_{N-2}+e^{-g}\psi _{0}  = \epsilon \psi _{N-1} \label{boundary}
\end{equation}%
and
\begin{equation}
e^{g}\psi_{N-1} +V_0\psi_{0}+e^{-g}\psi _{1} = \epsilon \psi _{0}. \label{BE2}
\end{equation}%
The homogeneous equation can be solved by first solving its characteristic equation
\begin{equation}
\epsilon = \frac{e^{g}}{z} +e^{-g}z. \label{character}
\end{equation}
Given an eigenenergy $\epsilon$, there exist two solutions $z_1$ and $z_2$ satisfying the constraint condition
\begin{equation}
\label{z1z2}
z_1 z_2=e^{2g},
\end{equation}
and thus the eigenenergies can be represented as $\epsilon = (z_1+z_2)e^{-g}$.
The general wavefunction takes the form of
\begin{equation} \label{solution}
\psi_{n} =\alpha_1z_1^{n}+\alpha_2z_2^{n}, 
\end{equation}
which fulfills the bulk eigen-equation of Eq. (\ref{eigeneq}). 
To obtain the eigensolutions of the whole system, the general ansatz of wave function in Eq.(\ref{solution}) shall also satisfy the boundary conditions.
Substituting Eq.(\ref{solution}) into Eq.(\ref{boundary}) and Eq.(\ref{BE2}), we obtain
\begin{equation} \label{AABB}
M_B \left(
\begin{array}{c}
\alpha_1\\
\alpha_2%
\end{array}%
\right)
=
\left(
\begin{array}{cc}
A(z_1,N)& A(z_2,N) \\
B(z_1,N)& B(z_2,N)%
\end{array}%
\right)
\left(
\begin{array}{c}
\alpha_1\\
\alpha_2%
\end{array}%
\right)=0
\end{equation}%
with
\begin{equation*}
A(z,N)=e^{-g}z+e^{g}z^{N-1}-\epsilon+V_0
\end{equation*}%
and
\begin{equation*}
B(z,N)=e^{g}z^{N-2}+e^{-g}-\epsilon z^{N-1}.
\end{equation*}%
\begin{figure}[tbp]
\includegraphics[width=0.45\textwidth]{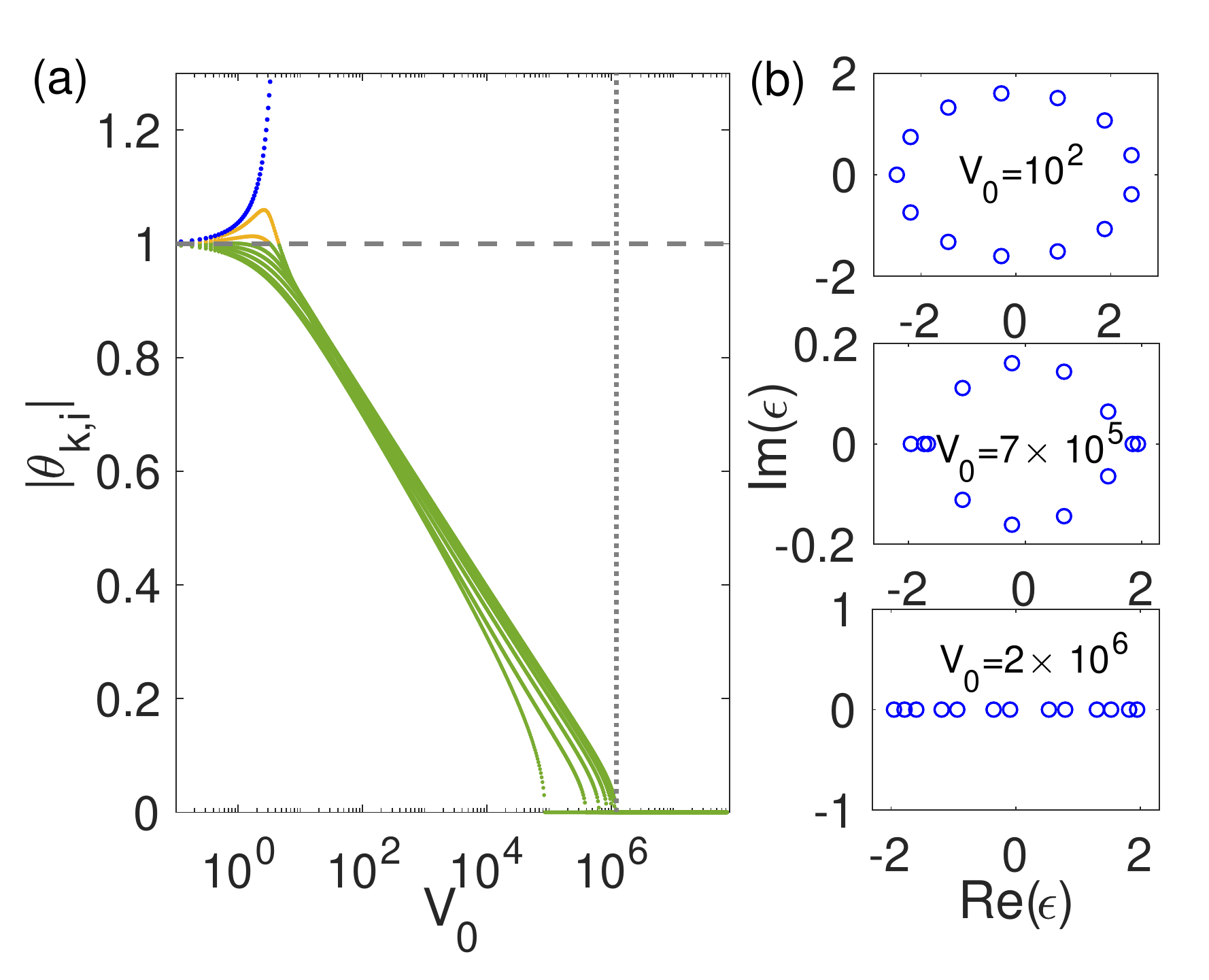}
\caption{(a) The absolute value of $\theta_{k,i}={\rm Im}[\theta_k]$ as a function of $V_0$. Parameters are $g=1$, $N=14$.
The blue dots are purely imaginary solutions ($\theta_{k,r}=0$) corresponding to 
the bound state localized at the impurity ($n=0$). 
Orange and green dots represent the solutions with $|\theta_{k,i}|>|g|$ and $|\theta_{k,i}|<|g|$ respectively.
The rest are the solutions corresponding to the bulk states.
The gray dashed line (horizontal) indicates $|\theta_{k,i}|=g=1$.
The gray dotted line (vertical) represents $V_0=2\sinh(gN)=1.2\times10^6$ for the chosen parameters.
(b) The corresponding complex spectra of bulk states for the system with 
$V_0=10^2$, $7\times 10^5$, and $2\times 10^6$, respectively. }
\label{fig2}
\end{figure}
The existence of nontrivial solutions for $(\alpha_1, \alpha_2)$ is determined by $\mathrm{det}[M_{B}] =0$,
which gives rise to the general solution with $\alpha_1\neq0, \alpha_2\neq0$:
\begin{equation}
\begin{split}\label{qqz1}
&(z_{1}^{N+1}-z_{2}^{N+1})-\frac{t _{R}}{t_{L}}%
(z_{1}^{N-1}-z_{2}^{N-1})\\
&-\left[ 1+\left( \frac{t_{R}}{t_{L}}\right) ^{N}\right] (z_{1}-z_{2})-\frac{V_0}{t_{L}}(z_{1}^{N}-z_{2}^{N})=0 .
\end{split}
\end{equation}
Eq.(\ref{qqz1}) and  Eq.(\ref{z1z2}) together determine the solution of $z_1$ and $z_2$ exactly.
Following the constraint condition of Eq.(\ref{z1z2}), we can always rewrite the solutions as
\begin{equation}\label{eq:z1_z2}
z_{1}=re^{i\theta}, ~~~ z_{2}=re^{-i\theta}
\end{equation}
with $r=\sqrt{\frac{t_{R}}{t_{L}}}=e^g$. 
Note that $\theta$ is not restricted to be real, which is important in determining the properties of the spectrum in latter discussion.
Thus Eq.(\ref{qqz1}) becomes
\begin{equation}\label{eq-theta}
2\cos[N\theta]\sin\theta-(r^{-N}+r^{N})\sin\theta- V_0 \sin[N\theta]=0,
\end{equation}
or equivalently
\begin{equation} \label{SE}
\frac{\sin\theta[2\cos(N\theta)-2\cosh(Ng)]}{\sin(N\theta)}=V_0 .
\end{equation}%
Defining $e^{i\theta}\equiv\beta$, Eq. \eqref{eq-theta} becomes a polynomial equation of $\beta$ with an order of $2N$, and hence shall have $2N$ different solutions. However, these solutions come in pairs with $\theta$ and $-\theta$, as the equation is invariant with the replacement $\beta\rightarrow 1/\beta$. Thus we obtain $N$ independent roots of $\theta=\theta_k$ with $k=1,2,...N$ labeling the $N$ roots, corresponding to $N$ eigenenergies 
\begin{eqnarray}
\epsilon_k=e^{-g}(z_1 +z_2)=2\cos\theta_k.\label{eq:epsilon}
\end{eqnarray}
We can see that the other $N$ roots $\theta=-\theta_k$ correspond to the same eigenenergies,
and the same eigenstates as the coefficients $\alpha_{1,2}$ in Eq. \eqref{solution} are also exchanged [which can be seen from Eqs. \eqref{AABB} and Eqs. \eqref{eq:z1_z2}].
For the PBC case with $V_0=0$, the solutions of Eq. \eqref{eq-theta} are given by 
$$\theta_k=\frac{2 k\pi}{N}+ig,~~k=1,2,\cdots,N.$$ The corresponding eigenenergies are complex 
and form an ellipse in the complex spectral plane. The eigenstates are given by
\begin{equation} \label{vzero}
\psi^k_{n} =\frac{1}{\sqrt{N}}e^{ng}e^{i n \theta_k}=\frac{1}{\sqrt{N}}e^{i 2 n k\pi /N},
\end{equation}
which are uniformly distributed. Notes that in this case $A(z_1,N)=0$, allowing us
to omit the branch $z_2$.

We then move on to the more sophisticated case where $V_0\neq 0$.
We first rewrite $\theta_{k}$ as $\theta_{k}=\theta_{k,r}+i\theta_{k,i}$,
where $\theta_{k,r}$ and $\theta_{k,i}$ are the real and imaginary part of $\theta_k$.
The energy $\epsilon_k$ is real either when $\theta_{k,i}=0$ or $\theta_{k,r}=(\pi\pm\pi)/2$, otherwise it acquires a nonzero imaginary amplitude.
With further analysis in appendix A, we find that $\theta_k$ is always real when $V_0$ exceeds a critical value $V_{0,c}$, with
$$V_{0,c}\approx2\sinh(Ng)\approx e^{Ng}$$
when $N\gg1$.
For relatively small $N$, the actual critical value $V_{0,c}$ is silghtly smaller than $2\sinh(Ng)$.
Fig. \ref{fig2}(a) shows the imaginary part $\theta_{k,i}$ for 
the system with $g=1$ and $N=14$ as a function of $V_0$,
with $V_0=2\sinh(Ng)$ given by the gray dotted line.
When $0<V_0\lesssim8.6\times10^4$, we obtain $|\theta_{k,i}|>0$ for every $k$, and the spectrum still forms an ellipse, as
shown in the top panel of Fig. \ref{fig2}(b). We note that the bound state is omitted and the eigenenergies of bulk states are demonstarted in Fig. \ref{fig2}(b). 
For $8.6\times10^4\gtrsim V_0<2\sinh(Ng)$, $\theta_{k,i}$ becomes zero for some values of $k$, 
meaning that a part of the spectrum becomes real, as shown in the middle panel of Fig. \ref{fig2}(b).
Due to the pseudo-Hermiticity, the non-real eigeneneriges in the above two cases always come in pairs with complex conjugated values of $\theta$, and hence the same $|\theta_{k,i}|$. 
When $V_0\geqslant2\sinh(Ng)$, all $\theta_{k,i}=0$ for bulk states, i.e. all $\theta_{k}$ and 
eigenenergies are real, as shown in the bottom panel of Fig. \ref{fig2}(b). 

\begin{figure}[tbp]
\includegraphics[width=0.45\textwidth]{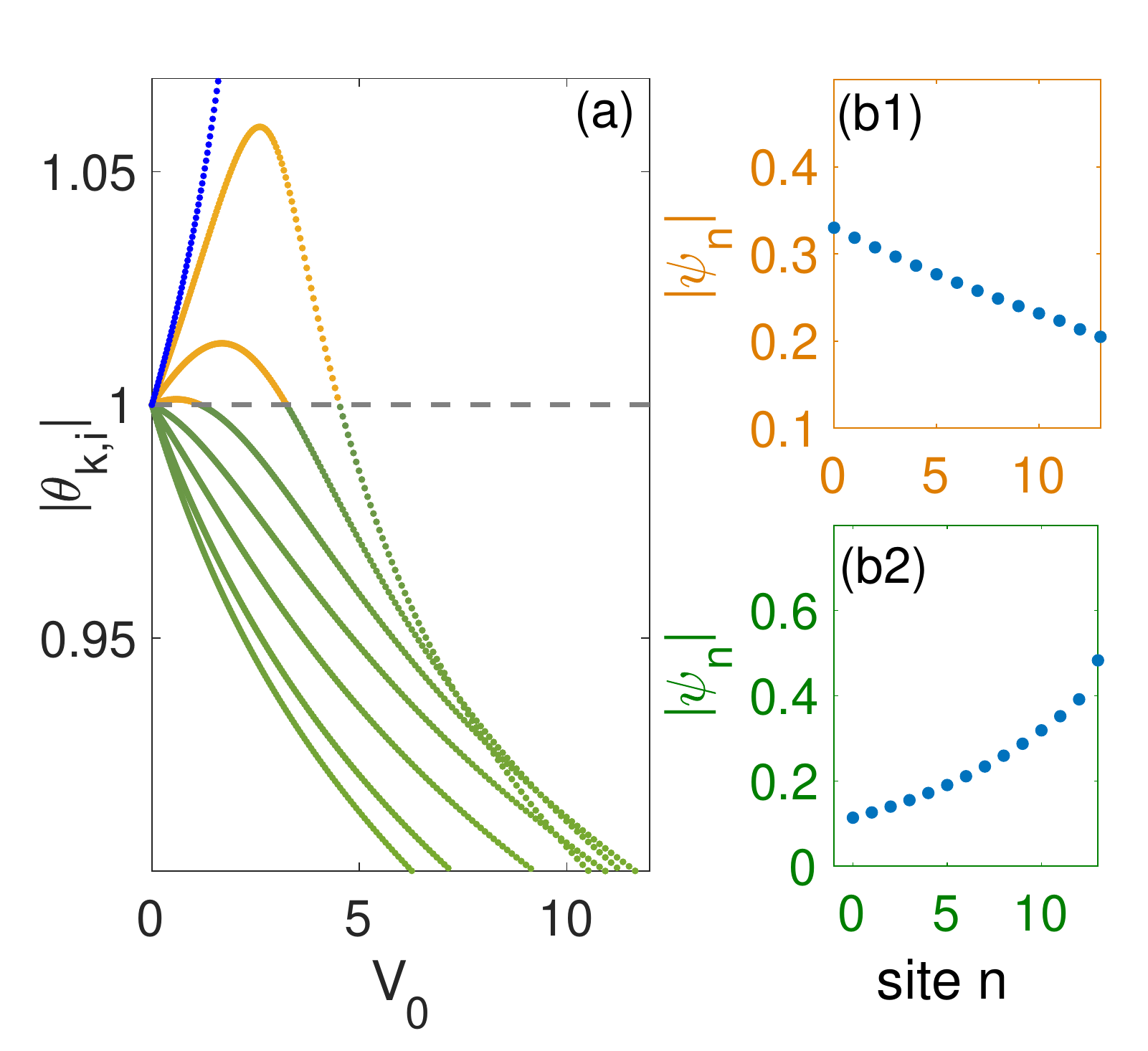}
\caption{(a)Zoom in of Fig. \ref{fig2} at regime of the weak $V_0$.
The gray dashed line indicates $|\theta_{k,i}|=g=1$.
(b) $|\psi_n|$ with different accumulating behaviors for (b1) $|\theta_{k,i}|>|g|$ and (b2) $|\theta_{k,i}|<|g|$.}
\label{fig3}
\end{figure}

Next we consider the behaviors of the eigenstates. 
When $V_0>0$ , the eigenstates \eqref{solution} 
can be written as
$$\psi_{n}^{k} = e^{ng}( \alpha_1e^{in\theta_k}+\alpha_2e^{-in\theta_k}).$$
 For weak impurity coupling $V_0$, $\theta_k$ are alway complex. We can see that the behavior
 of the wavefunctions depends not only on $\theta_k$, but also on $$\alpha_1/\alpha_2=-A(g,V_0,z_2,N)/A(g,V_0,z_1,N).$$ 
 It is straightforward to see that for $g>0$,
we shall have $\alpha_1/\alpha_2\gg1$ and $\psi_{n}^{k} \propto e^{n(g-\theta_{k,i})}e^{in\theta_{k,r}}$ when $\theta_{k,i}>0$, 
and $\alpha_2/\alpha_1\gg1$ and $\psi_{n}^{k} \propto e^{n(g-|\theta_{k,i}|)}e^{in\theta_{k,r}}$ when $\theta_{k,i}<0$.
Similar results can also be obtained for $g<0$. Overall, the wavefunction can be written as
\begin{equation} \label{wavefunction}
\psi_{n}^{k} \approx \alpha e^{\text{sgn}(g)(|g|-|\theta_{k,i}|)n}e^{in\theta_{k,r}}
\end{equation}
with $\alpha={\rm max}[\alpha_1,\alpha_2]$,
and the wavefunction decays with increasing $n$ when $|\theta_{k,i}|>|g|$, and with decreasing $n$ when $|\theta_{k,i}|<|g|$.
In Fig. \ref{fig2}(a), it is seen that $|\theta_{k,i}|>|g|$ is satisfied only in a regime with small $V_0$.
Therefore we zoom in and provide a clearer view of this regime in Fig. \ref{fig3}(a).
When $V_0\lesssim4.5$, some $\theta_{k,i}$ are seen to be greater than $|g|=1$, and 
the corresponding eigenstate $|\psi_n|$ decays to the right, as shown in Fig. \ref{fig3}(b1). 
When $V_0\gtrsim4.5$, all $\theta_{k,i}$ satisfy $|\theta_{k,i}|<g$, and the corresponding $|\psi_n|$ 
decays to the left, as shown in Fig. \ref{fig3}(b2).

\begin{figure}[tbp]
\includegraphics[width=0.48\textwidth]{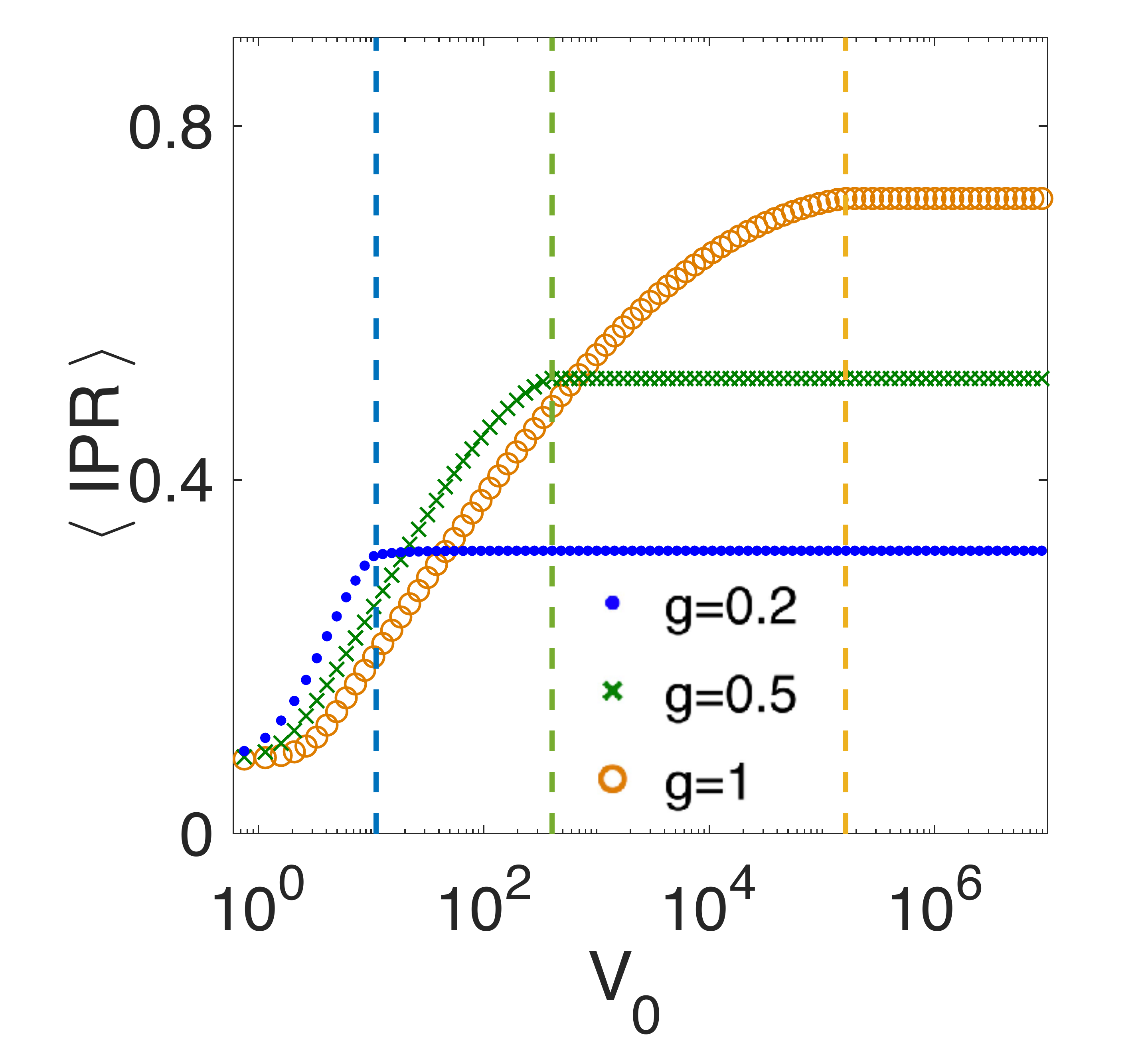}
\caption{Average IPR for the system (\ref{H0})  with $N=14$, $g=0.2$, $0.5$, and $1$, respectively.
The dashed lines represent the transition points from non-skin to skin states: $V_0=2\cosh(Ng)$.}
\label{fig4}
\end{figure}

When $V_0\geqslant2\sinh(Ng)$, $\theta_{k,i}=0$ is obtained for all bulk states, and the wavefunction 
$\psi_{n}^{k} \propto e^{gn}e^{in\theta_{k,r}}$ takes the form of skin states.
That is, all bulk states now become skin states and all the eigenenergies are real,
which is the same situation as the system with $V_0=0$ but under OBC.
To see clearly the transition from the non-skin
states to skin states, we calculate the averaged inverse
participation ratio (IPR) to characterize the localization of the system,
\begin{equation*}
\langle \text{IPR} \rangle =\frac{1}{N}\sum_k \frac{\sum_{n}\left\vert \left\langle n|\psi^k \right\rangle
\right\vert ^{4}}{\left( \left\langle \psi^k |\psi^k \right\rangle \right) ^{2}},
\end{equation*}%
where $\left\vert \psi^k \right\rangle $ is the $k$-th right eigenstate of Hamiltonian (\ref{H0}).
The non-skin states given by Eq. \eqref{wavefunction} are sensitive to $V_0$ and size $N$, suggesting that variation of $V_0$ or $N$ can
change $\langle \text{IPR} \rangle$ significantly, analogous to the scale-free accumulating states in Ref. [\onlinecite{Linhu}]. On the other hand, the locality of skin states only depends on $g$ and shall give a constant $\langle \text{IPR} \rangle$ independent from the impurity strength $V_0$.
We display the $\langle \text{IPR} \rangle$ for all bulk states versus $V_0$ in Fig. \ref{fig4}, which
indicates a clear transition from non-skin to skin states at
$V_0=2\cosh(Ng)$. That is, as $V_0$ is increased, $\langle \text{IPR} \rangle$ increases before the transition point of $V_0=V_{0,c}$,
and remains a constant for the skin states after the transition.

\begin{figure}[tbp]
\includegraphics[width=0.48\textwidth]{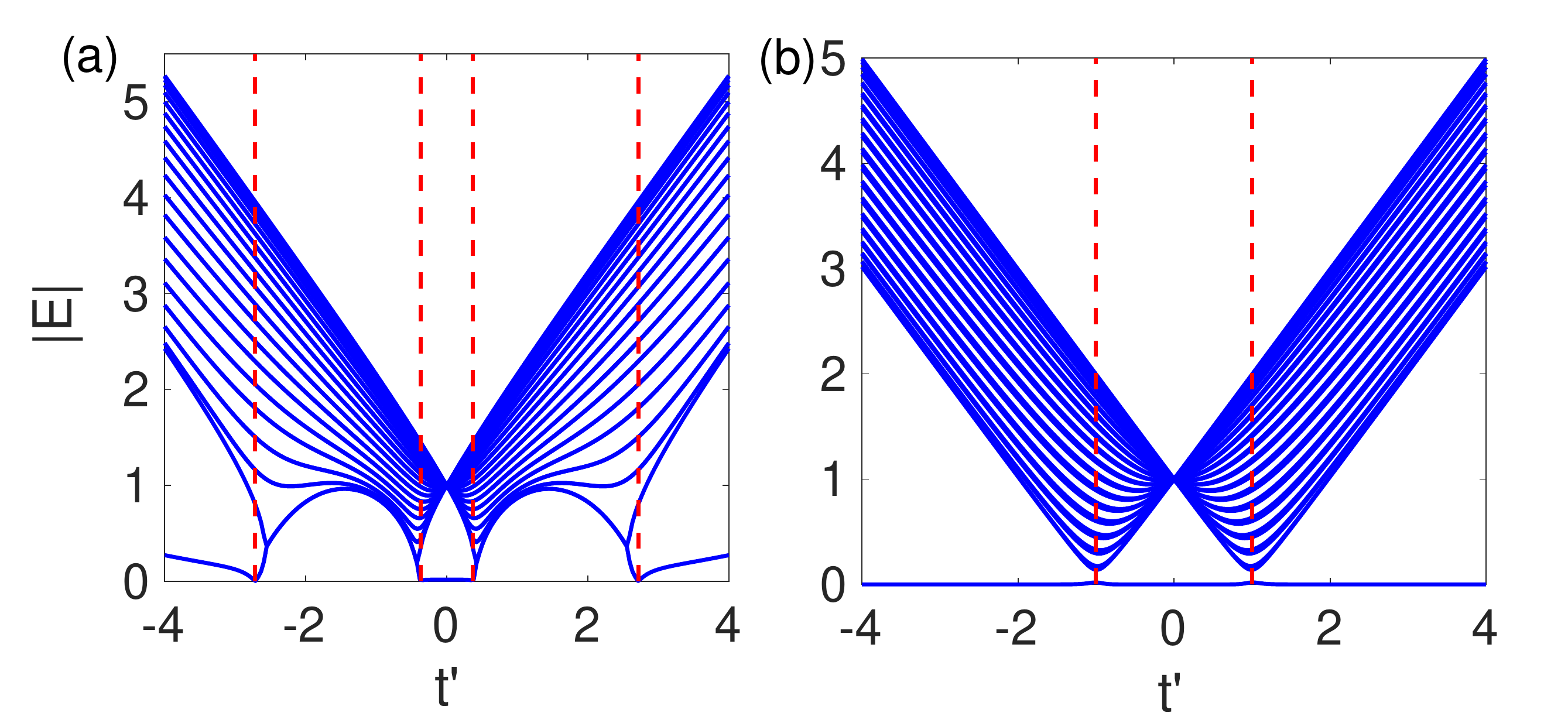}
\caption{ Spectrum $|E|$ for the system  (\ref{impSSH})  with $N=20$, $g=1$,
(a) $V=2 \cosh(4)$ and (b) $V=2 \cosh(N+1) $, respectively.
The dashed lines in (a) represent the gap-closing points $t'=\pm e^{\pm g}$. 
The dashed lines in (b) represent the gap-closing points $t'=\pm 1$.}
\label{fig5}
\end{figure}

\section{Single impurity in non-reciprocal SSH Model}

The non-reciprocal SSH model \cite{SSH} with a single impurity under PBC, as 
displayed in Fig. \ref{fig1}(b), can also be exactly solved. 
The Hamiltonian is given by
\begin{eqnarray}                 \label{impSSH}
H_{\rm SSH}&=&\sum_{n=0}^{N-1}\left[ t (e^{-g}\left\vert n,A\right\rangle
\left\langle n,B\right\vert +e^{g}\left\vert n,B\right\rangle \left\langle
n,A\right\vert ) \right.  \notag \\
&& \left. + t^{\prime } ( \left\vert n,B\right\rangle \left\langle
n+1,A\right\vert + h.c.)\right]+V_0 \left\vert 0,A \right\rangle  \left\langle
0,A \right\vert, \notag \\             
\end{eqnarray}
with A and B indexes for different sublattices, 
$\left\vert N,A(B) \right\rangle \equiv \left\vert 0,A(B) \right\rangle$,
$N$ the total number of unit cells, 
$V_0$ the impurity strength,
and $t'$ ($te^{\pm g}$) the inter-cell (right and left intra-cell) hopping term(s).
The energy unit is set to be $t=1$ for latter convenience. 

When $V_0=0$, through the Fourier transformation, the Hamiltonian of the non-reciprocal SSH model without the impurity
can be represented as
\begin{equation*}
H_{0}\left( k\right) =h_{x}\sigma _{x}+h_{y}\sigma _{y}
\end{equation*}%
in the momentum space, where $h_{x}=t^{\prime }\cos k+\cosh g$, $h_{y}=t^{\prime }\sin k-i\sinh g$
with $0\leqslant k<2\pi$,
and 
$\sigma _{x,y}$ are the Pauli matrices. 
The eigenenergies are given by
\begin{eqnarray*}
E&=&\pm\sqrt{(h_x+i h_y)(h_x-i h_y)}\\
&=&\pm\sqrt{(t^{\prime}e^{i k}+e^g)(t^{\prime}e^{-i k}-e^{-g})}.
\end{eqnarray*}%

The two-band structure of this model allows line-gap topology to emerge in this system, 
and the phase boundaries can be determined by the gap closing conditions of the system \cite{Yin,Yao},
\begin{equation}
t^{\prime }=-e^{\pm g}~~~\text{and}~~~t^{\prime }=e^{\pm g}. \label{condition1}
\end{equation}%
Note that through the similarity transformation, the model \eqref{impSSH} with $V_0=0$ under OBC 
becomes the Hermitian SSH model, where 
the topological transition points (gap-closing points) are given by $t'=\pm t=\pm1$. 
In other words, the spectra of systems under PBC and OBC have different gap-closing points \cite{guo2021exact,Yao,Kunst}

When $V_0\neq 0$, we can solve Eq. \eqref{impSSH} by taking the wavefunction as 
$\left\vert \psi \right\rangle =\sum_{n}\left(\psi_{A,n}\left\vert n,A\right\rangle +\psi _{B,n}\right)
 \left\vert n,B\right\rangle $. The stationary Schr\"{o}dinger equation 
$H_{\rm SSH}\left\vert \psi \right\rangle =E\left\vert \psi
\right\rangle $ is equivalent to the
following difference equations%
\begin{eqnarray}
e^{g}\psi _{A,n-1} +t^{\prime }\psi _{A,n}
=E \psi _{B,n-1} ,  \label{re1} \\
t^{\prime }\psi _{B,n-1} +e^{-g}\psi _{B,n} +\delta
_{x,0}V_0\psi _{A,n} =E \psi _{A,n},
\label{re2}
\end{eqnarray}%
where $E$ is the eigenenergy. Eq.(\ref{re1}) gives us
\begin{equation}
\psi _{B,n} = \frac{e^{g}}{E }\psi _{A,n} + \frac{t^{\prime }}{E }\psi _{A,n+1}.
\label{re3}
\end{equation}%
We can decouple $\psi _{A}$ and $\psi _{B}$ by substituting Eq.(\ref{re3}) into Eq.(\ref{re2}), which yields %
\begin{eqnarray}
& & e^{g}\psi _{A,n-1} +e^{-g}\psi _{A,n+1}
+\delta _{n,0}\frac{E V_0}{t^{\prime }}\psi _{A,n}
\notag \\
&=& \frac{E ^{2}-t^{\prime 2}-1}{t^{\prime }}\psi _{A,n}.  \label{Eq-A-lattice}
\end{eqnarray}
It is obvious that Eq. \eqref{Eq-A-lattice} under the following substitution
\begin{eqnarray}
& &V_0 \rightarrow \frac{E V_0}{t^{\prime }},  \label{veff} \\
& &\epsilon \rightarrow \frac{E ^{2}-t^{\prime 2}-1}{t^{\prime }}.
\label{Eeff}
\end{eqnarray}
is identical to Eq. \eqref{eigeneq} with boundary 
conditions (\ref{boundary}) and \eqref{BE2}.
Thus to solve Eq. \eqref{Eq-A-lattice}, we can directly use the results of the
impurity problem in the HN model in section II.
That is, by substituting Eqs. \eqref{Eeff} and \eqref{veff} into Eqs. \eqref{eq:epsilon}) and \eqref{SE},
we obtain
$$E=\pm \sqrt{1+t^{\prime 2} +2t^{\prime}\cos\theta}$$
and
\begin{equation} \label{SEssh}
\frac{t'\sin(\theta)(2\cos(N\theta)-2\cosh(Ng))}{\pm \sqrt{1+t^{\prime 2} +2t^{\prime}\cos\theta}\sin(N\theta)}=V_0
\end{equation}
respectively.
Following our analysis of Eq. \eqref{SE}, the transition point from non-skin states
to skin states can be determined by $$V_{0,c}=2 t^* \sinh(Ng)\approx t^*e^{Ng}$$ with $t^*=\min\{1,t'\}$ for large $N$ or $g$. 
$V_{0,c}$ increases exponentially with the lattice size $N$.
At small $V_{0,c}$ far from the critical value,
all $\theta_k$ for bulk states 
are complex. The  spectra are similar to the system with $V_0=0$. The topological phase 
transition occurs at $t'=\pm \text{e}^{\pm g}$, as shown in Fig. \ref{fig5} (a).
When $V_0\geqslant2 t^* \sinh(Ng)$, all $\theta_k$ for bulk states are real.
So all the bulk states become the skin states and all the eigenenergies are real,
analogous to the system with $V_0=0$ under OBC. The topological phase 
transition then takes place at $t'=\pm 1$, as shown in Fig. \ref{fig5} (b).

\begin{figure}[tbp]
\includegraphics[width=0.48\textwidth]{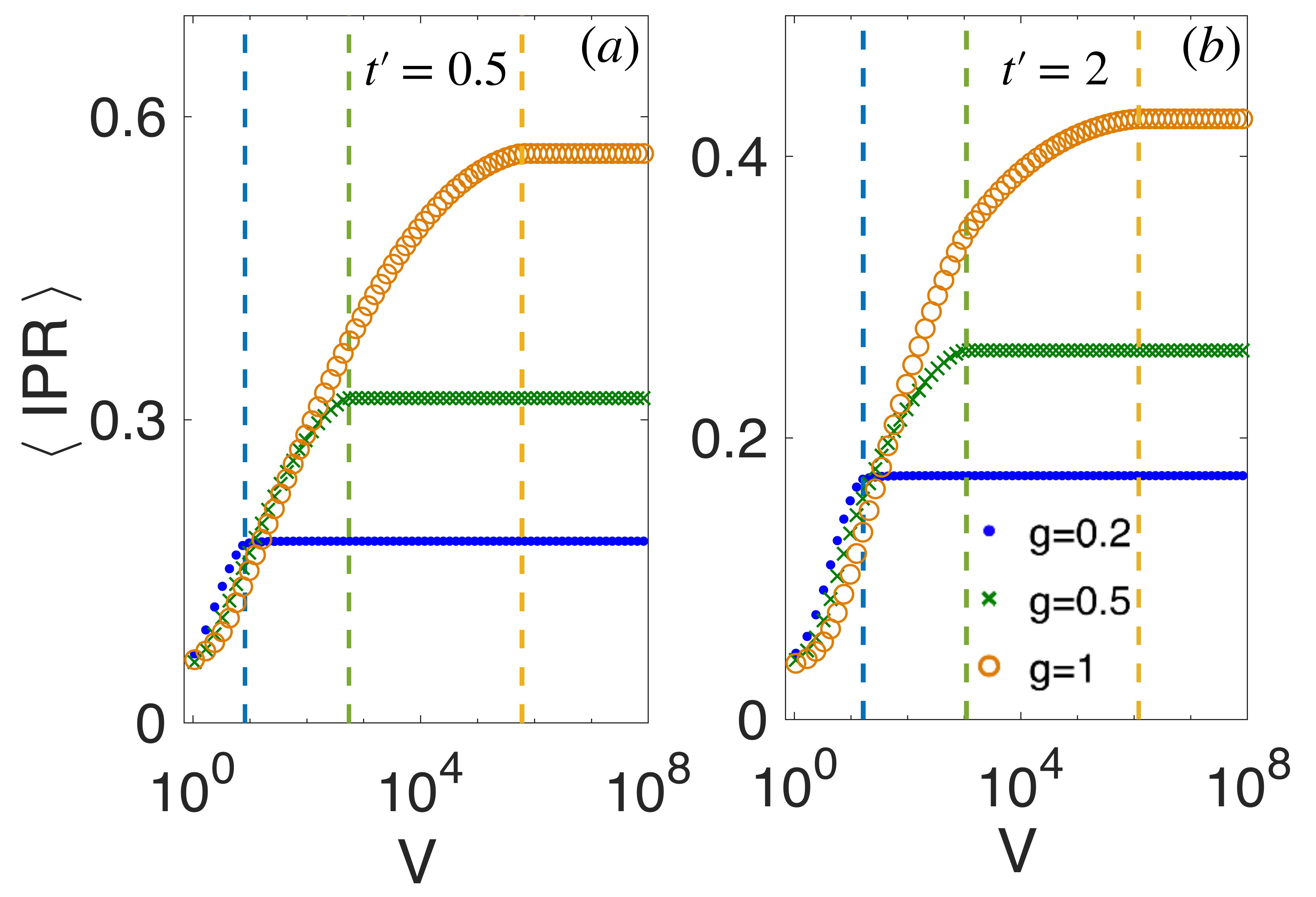}
\caption{(a) Average IPR for the system (\ref{impSSH})  with $N=14$ $t'=0.5$, $g=0.2$, $0.5$, and $1$, respectively.
The dashed lines represent the transition points from non-skin to skin states, $V_{0,c}=\sinh(Ng)$.
(b) Average IPR for the system (\ref{impSSH})  with $N=14$ $t'=2$, $g=0.2$, $0.5$, and $1$, respectively.
The dashed lines represent the transition points from non-skin to skin states, $V_{0,c}=2\sinh(Ng)$.}
\label{fig6}
\end{figure}

Finally, we display $\langle \text{IPR} \rangle$ of all bulk states versus $V_0$ in Fig. \ref{fig6} with $t'=0.5$ 
and $t'=2$, respectively. Fig. \ref{fig6} demonstrates a clear transition from non-skin states to skin states at
$V_0=\sinh(Ng)$ with $t'=0.5$ and $V_0=2\sinh(Ng)$ with $t'=2$ for $N=14$.  
$\langle \text{IPR} \rangle$ increase for the non-skin states with increasing $V_0$,
and remains constant for the skin states.

\section{Quantized response of the spectral topology in the impurity models}
In previous sections, we have exactly solved the single-impurity problem for the HN model and SSH model, both exhibit the NHSE and thus possess a non-trivial spectral topology \cite{Borgnia,KZhang,Okuma}.
It has recently been discovered that a quantized response corresponding to the spectral topology can be extracted from the system's Green's function element, in the process of tuning the boundary condition from PBC to OBC continuously \cite{li2020quantized}.
Therefore we expect a similar phenomenon to arise also in our model, as a strong impurity strength effectively acts as the OBC.

Following Ref. [\onlinecite{li2020quantized}], we define quantities as
\begin{eqnarray}
\nu_{0(N-1)}=\frac{\partial \ln G_{0(N-1)}}{\partial \ln V_0},
~~\nu^B_{0(N-1)}=\frac{\partial \ln G^B_{0(N-1)}}{\partial \ln V_0}
\end{eqnarray}
for the HN model and the SSH model respectively, with $$G_{0(N-1)}=\left\langle 0 \right\vert  G_{\rm HN}\left\vert N-1 \right\rangle$$
and $$G^B_{0(N-1)}=\left\langle 0,B \right\vert  G_{\rm SSH}\left\vert N-1,B \right\rangle$$
the off-diagonal elements of the Green's functions $G_{\rm HN}$ of the HN model and $G_{\rm SSH}$ of the SSH model,
\begin{eqnarray}
G_{\rm HN}=(E_r-H_{\rm HN})^{-1}, G_{\rm SSH}=(E_r-H_{\rm SSH})^{-1}.
\end{eqnarray}
$E_r$ is a chosen complex reference energy to define the spectral winding \cite{KZhang,Okuma,li2020quantized}.
In our models, the spectral winding number is $\nu=1$ for $E_r$ enclosed by the PBC spectra in the complex plane. In such cases, $\nu_{0(N-1)}$ and $\nu^B_{0(N-1)}$ are expected to jump from $1$ to $0$ for the concerned models when $V_0$ is increased and reaches a critical value, where the spectra coincide with $E_r$ \cite{li2020quantized}.

\begin{figure}[tbp]
\includegraphics[width=0.48\textwidth]{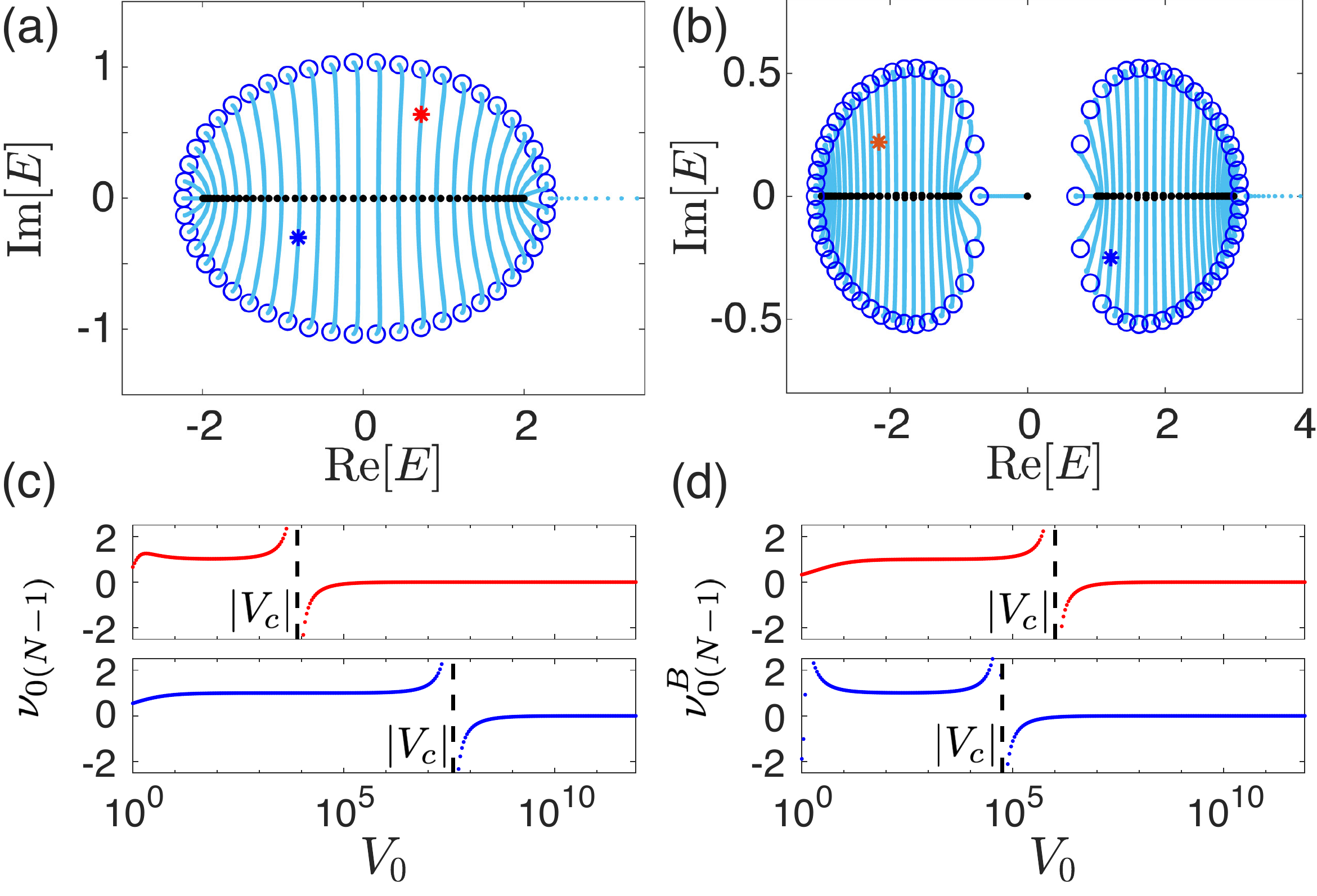}
\caption{(a) (b) The spectra with $V_0=0$ (blue circles) and $V_0=10^{12}$ (black dots) for the HN model and the SSH model respectively. Cyan dotted-curves illustrate the spectra with $V_0$ varying from $0$ to $10^{12}$, namely the spectral flow from PBC to strong impurity limit, the latter effectively gives the OBC. 
Red and blue stars are the chosen reference energies $E_r$ for calculating $\nu_{0(N-1)}$ and $\nu_{0(N-1)}^B$, $E_r=0.72+0.64i$ (red) and $-0.81-0.3i$ (blue) in (a), and $E_r=-2.16+0.22i$ (red) and $1.21-0.25i$ in (b).
(c) (d) The defined quantities $\nu_{0(N-1)}$ and $\nu_{0(N-1)}^B$ versus $V_0$ for different $E_r$, indicated by the stars with the same colors in (a) and (b). The black dash lines indicate the critical value $|V_c|$ obtained from Eqs. \eqref{SE} and \eqref{SEssh} by requiring $\epsilon=E_r$. Note that the obtained $V_c$ takes complex values as $E_r$ does not exactly fall on the spectral flow.
}
\label{fig7}
\end{figure}

In Fig. \ref{fig7}, we illustrate the defined quantities as functions of $V_0$ for several different $E_r$ enclosed by the PBC spectrum at $V_0=0$.
It is seen that each of $\nu_{0(N-1)}$ and $\nu_{0(N-1)}^B$ roughly exhibits a plateau at $1$, and jumps to $0$ after a critical value $V_0=V_c$. 
Note that $V_c$ is distinguished from $V_{0,c}$ in previous sections, which describes the transition to a fully real spectrum. 
The spectrum forms a shrinking ellipse with increasing $V_0$, passing through $E_r$ at $V_0=V_c$, after which the spectral winding jumps from $1$ to $0$.
Therefore for a given $E_r$, we can require it to be an eigenenergy $\epsilon$ of the system, then determine $V_c$ through Eqs. \eqref{SE} and \eqref{SEssh} for the two models respectively. 

We would also like to point out that in a finite-size system, the spectral flow [cyan dotted curves in Fig. \ref{fig7}(a) and (b)] from $V_0=0$ to $V_0\rightarrow \infty$ cannot cover the regime enclosed by the PBC ellipse-spectrum completely. Strictly speaking, a real $V_c$ can be obtained only when $E_r$ exactly falls along the spectral flow. Nevertheless, numerically we can choose $E_r$ close enough to the spectral flow [blue and red stars in Fig. \ref{fig7}(a) and (b)], and the absolute value of the obtained $V_c$ are seen to be in good consistency with the jump of $\nu_{0(N-1)}$ and $\nu_{0(N-1)}^B$ in Fig. \ref{fig7}(c) and (d) respectively.

\section{Summary}
In summary, we have exactly solved the impurity problem in the HN model and the SSH model. 
The exact solutions of finite-size systems reveal a transition for the bulk states from non-skin states to skin states when increasing the impurity strength $V_0$,
and the corresponding complex eigenenergies also become real after the transition.
The critical value $V_{0,c}$ of the impurity for the transition depends on both the lattice size $N$ and the parameter $g$ describing the non-reciprocaity, 
and increases as $V_{0,c}=\sinh{Ng}$. $\sinh{Ng}$ serves as exact $V_{0,c}$ in the large $N$, while still of high accuracy in the small $N$, despite
the value of $g$.
Such a transition indicates that a strong impurity acts as an open boundary for the NHSE in non-reciprocal non-Hermitian systems.
Different bulk states are also found to reach the OBC limit at different critical values of $V_{0,c}$. 
We have also extended our study to the single-impurity problem of the SSH model, which can be mapped to the HN model, and exact solutions can be obtained accordingly.
Our exact solutions are also proven useful for investigating the spectral topological transition in the concerned models.

\begin{acknowledgments}
The work is supported by NSFC under Grants No.11974413 and the National Key
Research and Development Program of China (2016YFA0300600 and
2016YFA0302104).
L. L. acknowledges funding support by the Guangdong Basic and Applied Basic Research Foundation (No. 2020A1515110773).
\end{acknowledgments}

  \appendix
  \begin{figure*}[tbp]
\includegraphics[width=0.8\textwidth]{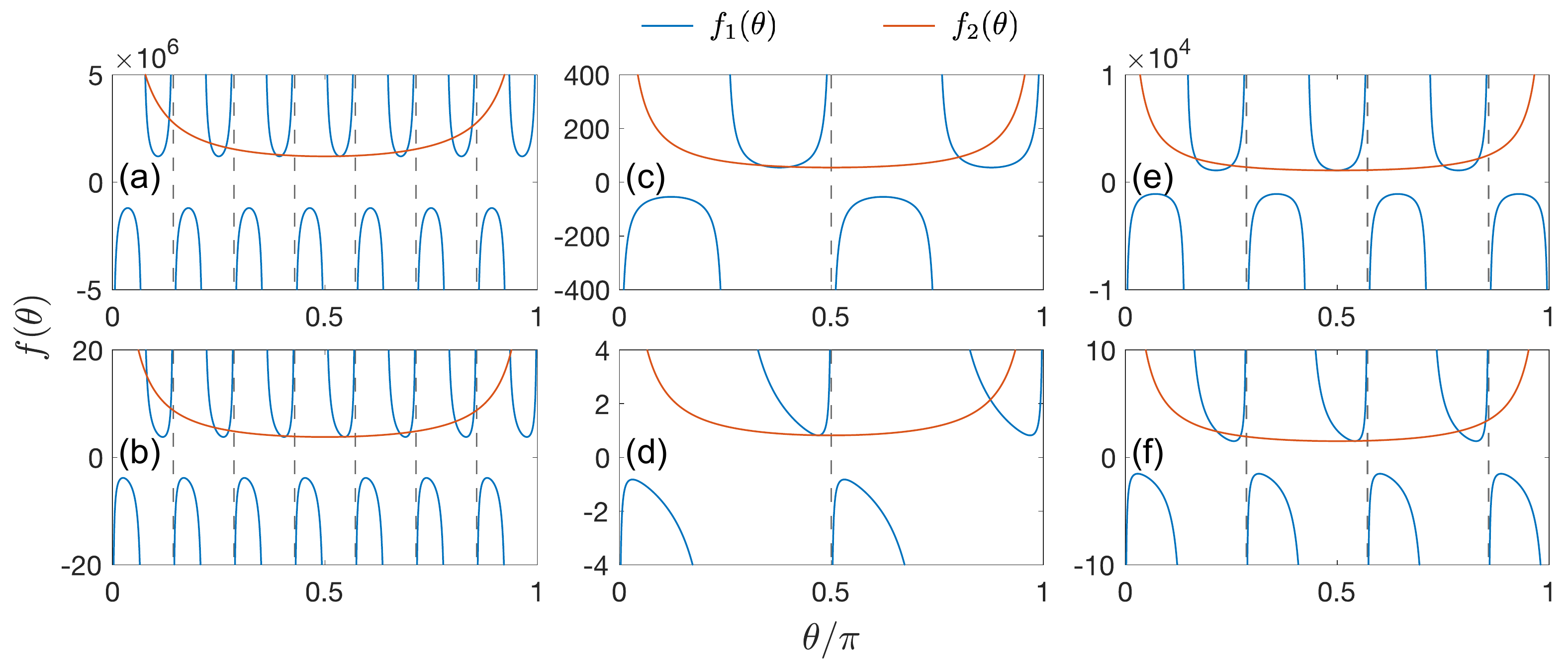}
\caption{ $f_1(\theta)$ and  $f_2(\theta)$ with $V_0=2\sinh [Ng]$ as function of $\theta$. Parameters given in (a) $N=14$, $g=1$;
(b) $N=14$, $g=0.1$; (c) $N=4$, $g=1$; (d) $N=4$, $g=0.1$; (e) $N=7$, $g=1$; and (f) $N=7$, $g=0.1$. Dash lines separate different period of $f_1(\theta)$.}
\label{Afig1}
\end{figure*}
\section{The transition point between complex and real spectra}
From Fig. \ref{fig2} we see that the spectrum of the HN model with an impurity becomes purely real when $V_0$ is large enough. To identify the critical value $V_{0,c}$ for this transition,
we rewrite Eq. \eqref{eq-theta} as
\begin{equation}\label{re-eq-theta}
f_1(\theta)=f_2(\theta),
\end{equation}
where 
\begin{equation}\label{f1}
\begin{split}
f_1(\theta)&=\frac{2\cos[N\theta]-(r^{-N}+r^{N})}{\sin[N\theta]} \\
& =\frac{2\cos[N\theta]-2\cosh(Ng)}{\sin[N\theta]},
\end{split}
\end{equation}
and 
\begin{equation}\label{f2}
f_2(\theta)=\frac{V_0}{\sin\theta}.
\end{equation}
Note that the energy $$\epsilon=2\cos\theta$$ is real with either purely real or imaginary $\theta$. 
We shall first focus on the case with real $\theta$, which has a period of $2\pi$. 
On the other hand, $\theta$ and $-\theta$ correspond to the same eigensolution of the system, as discussed in the main text.
Therefore we only need to focus on the behavior of $f_{1,2}(\theta)$ with $\theta\in(0,\pi)$.
Here the two points of $\theta=0$ and $\pi$ are excluded, because at such values we shall 
obtain a non-physical solution with a vanishing wavefunction $\phi_n=0$ by substituting Eq. \eqref{eq:z1_z2} into Eq. \eqref{AABB} in the main text.
If $f_1(\theta)$ and $f_2(\theta)$
have $N-x$ intersection points, Eq. \eqref{re-eq-theta} has $N-x$ real
and $x$ complex solutions, where $0\leqslant x \leqslant N$.
In this regime, $f_2(\theta)$ is a positive definite function, which is symmetrical of $\theta=\pi/2$ and its 
minimum value is given by $f_{2,min}=f_2(\pi/2)=V_0$.
$f_1(\theta)$ has a period of $2\pi/N$ (separated by the dash lines in Fig. \ref{Afig1}), and satisfies $f_1(\theta)>0$ in the regimes of 
\begin{eqnarray}\label{interval}
\theta \in \left(\frac{(2n+1)\pi}{N},\frac{(2n+2)\pi}{N}\right),~~n=0,1,\cdots,\lfloor \frac{N}{2}\rfloor-1,\nonumber\\
\end{eqnarray}
i.e. the right half of each period.
In each interval, $f_1(\theta)$ decreases monotonically at first and meets its minimum value $f_{1,min}=2\sinh{Ng}$
at 
\begin{eqnarray}
\theta=\frac{1}{N}(-\arccos \frac{1}{\cosh (Ng)}+ 2n\pi),~~n=1,\cdots,[N/2],\nonumber
\end{eqnarray}
and increases monotonically again (see Fig. \ref{Afig1}).
Therefore $f_1(\theta)$ and $f_2(\theta)$ generally intersect twice within each of the $\lfloor N/2\rfloor$ intervals given by Eq. \eqref{interval}. However, for an even $N$, the second intersection in the last interval tends to $\theta=\pi$ where both $f_1(\theta)$ and $f_2(\theta)$ tend to infinity.
Thus we obtain $N-1$ intersections of $f_1(\theta)$ and $f_2(\theta)$ in total for $\theta\in(0,\pi)$, with either an even [Fig. \ref{Afig1}(a)-(d)] or an odd $N$ [Fig. \ref{Afig1}(e) and (f)].
When $N \rightarrow \infty$, the period of $f_1(\theta)$ approaches zero, meaning we can find the minimum value of $f_1(\theta)$ infinitely
close to $\pi/2$, leading to
$$\lim_{N \rightarrow \infty }V_{0,c}=2\sinh (Ng),$$
and two intersections around this nadir emerge when $V_{0,c}>2\sinh (Ng)$.
As we can infer from Fig. \ref{Afig1}(c) and (d), $2\sinh(Ng)$ makes a good approximation of $V_{0,c}$ also for small $N$.
The exact critical value of $V_{0,c}$ can only be smaller than $2\sinh(Ng)$, as $f_2(\theta)$ gets larger when $\theta$ diverges away from $\pi/2$, and hence approaches the minimums of $f_1(\theta)$ at a smaller $V_0$.
To conclude, when $V_0\geqslant 2\sinh(Ng)$, $f_2(\theta)$ and $f_1(\theta)$ shall have $N-1$ intersections for $\theta\in(0,\pi)$, 
corresponding to $N-1$ real solutions of the bulk states.

For imaginary $\theta$, $f_1(\theta)$ and $f_2(\theta)$ are mostly monotonic and only 
gives a single intersection for ${\rm Im}[\theta]\in(0,\infty)$, regardless of the value of $N$ and $g$, 
corresponding to the bound state.

\end{document}